\documentclass[12pt,a4paper]{article}
\usepackage[latin1]{inputenc}
\usepackage{amsmath}
\usepackage{amsfonts}
\usepackage{amssymb}
\newcommand{\be}{\begin{equation}}
\newcommand{\ee}{\end{equation}}

\begin{document}

\begin{center}
{\Large\bf Relation between quark and gluon condensates from QCD sum rules}
\end{center}

\begin{center}
{\large S.S.~Afonin} \\
Departament d'ECM, 
Universitat de Barcelona, \\
Barcelona 08028, Spain\\
e-mail: afonin@ecm.ub.es
\end{center}

\begin{abstract}
A new relation between $\rho$-meson mass, weak $\pi$-meson decay constant, quark and gluon condensates is derived from the QCD sum rules. As a byproduct an explanation for the dominance of $\rho\rho$-decay for the $f_0(1370)$-meson is proposed. 
\end{abstract}

\noindent
Pacs: 12.38.Lg 

\section{Introduction}

Since the invention of SVZ sum rules approach~\cite{svz} the quark $\langle\bar{q}q\rangle$ and gluon $\langle G^2\rangle$ condensates have been usually taken as independent input parameters characterising the QCD vacuum. For the heavy quarks, however, there is 
relation~\cite{svz} 
\begin{equation}
\langle\bar{Q}Q\rangle=-\frac{1}{12m_Q}\frac{\alpha_s}{\pi}\langle G^2\rangle+\mathcal{O}(m_Q^{-3}).
\end{equation}
In the sector of light quarks, which we are interested in, such a direct expansion does not exist. The only
relation which we have is the low-energy theorem~\cite{nsvz}
\begin{equation}
\frac{d}{dm_q}\frac{\alpha_s}{\pi}\langle G^2\rangle=-\frac{24\langle\bar{q}q\rangle}{\frac{11}{3}N_c-\frac{2}{3}N_f}.
\end{equation}
This formula does not permit to calculate the value, say, of gluon condensate given the value of quark one.
An attempt to relate them was undertaken in~\cite{bec} (that analysis, however, resulted in a too big estimation for the values of current quark masses). 
In the present Letter we return to this problem and show that the QCD sum rules in the limit of large-$N_c$~\cite{hoof} and in the leading order of perturbation theory indeed give a certain relation between these condensates at some assumptions about the meson spectra.

The paper is organized as follows. Sect.~2 contains some general formulas. In Sect.~3 the derivation of Weinberg relation is sketched and a similar logic is applied to the scalar and pseudoscalar channels. Then the consequences of combining these results are considered.

\section{Two-point correlators in the large-$N_c$ limit}

In the limit of infinite number of colours the meson spectrum of QCD consists of infinite number of weakly interacting stable resonances. They saturate completely the two-point correlators of quark currents. Thus the $SU_f(2)$ two-point correlators of the scalar (S), pseudoscalar (P), vector (V), and axial-vector (A) quark currents can be represented as the following infinite sums in the Euclidean space
\begin{equation}
\label{cor1}
\Pi^J(Q^2)=\int d^4x\,e^{iQx}\langle\bar{q}\Gamma_J q(x)\bar{q}\Gamma_J q(0)
\rangle=
\sum_{n=0}^{\infty}\frac{Z_J(n)}{Q^2+m_J^2(n)}+D_0^J+D_1^JQ^2,
\end{equation}
\begin{equation}
J\equiv S,P,V,A; \qquad
\Gamma_J=i,\gamma_{5}\tau^a,\gamma_{\mu}\tau^a,\gamma_{\mu}\gamma_{5}\tau^a; \qquad
D_{0}^J,D_{1}^J=\text{const}.
\end{equation}
Here $D_{0}^J$ and $D_{1}^JQ^2$ represent contact terms required for the renormalization of infinite sums and $\tau^a$ are the standard Pauli matrices.
On the other hand, the Operator Product Expansion (OPE) for the same correlators in the large-$N_c$ and chiral limits looks as follows in the leading order of perturbation theory~\cite{rry,jam}
\begin{equation}
\label{V}
\Pi^{V,A}(Q^2)=-\frac{N_c}{12\pi^2}Q^2
\ln\!\frac{\Lambda^2}{Q^2}
-\frac{\alpha_s}{12\pi}
\frac{\langle G^2\rangle}{Q^2}+\mathcal{O}\left(\frac{1}{Q^4}\right),
\end{equation}
\begin{equation}
\label{S}
\Pi^{S,P}(Q^2)=-\frac{N_c}{8\pi^2}Q^2
\ln\!\frac{\Lambda^2}{Q^2}
+\frac{\alpha_s}{8\pi}
\frac{\langle G^2\rangle}{Q^2}+\mathcal{O}\left(\frac{1}{Q^4}\right),
\end{equation}
where we have defined
\begin{equation}
\label{trans}
\Pi_{\mu\nu}^{V,A}(Q^2)\equiv\left(-\delta_{\mu\nu}+\frac{Q_{\mu}Q_{\nu}}{Q^2}\right)
\Pi^{V,A}(Q^2).
\end{equation}
In the above expressions $\Lambda$ represents the momentum cut-off. For the present analysis we do not need the next condensate terms. 

The residues are parametrized as follows
\begin{equation}
\label{par}
Z_{V,A}(n)\equiv2F_{V,A}^2(n)m_{V,A}^2(n), \qquad Z_{S,P}(n)\equiv2G_{S,P}^2(n)m_{S,P}^2(n),
\end{equation}
with $F_{V,A}(n)$ and $G_{S,P}(n)$ being some decay constants parametrizing the corresponding matrix elements:
\be
\langle0|\bar{q}\Gamma_J q(0)|J(n)\rangle=\sqrt{2}F_J(n)m_J(n)\epsilon_{\mu},\qquad J=V,A,
\ee
\be
\langle0|\bar{q}\Gamma_J q(0)|J(n)\rangle=\sqrt{2}G_J(n)m_J(n),\qquad J=S,P.
\ee 
In the pseudoscalar case the $\pi$-meson pole is not subjected to parametrization~\eqref{par}, giving the contribution
\be
\Pi^P(Q^2)\longrightarrow\Pi^P(Q^2)+\frac{2}{Q^2}\frac{\langle\bar{q}q\rangle^2}{F_{\pi}^2},
\ee
where $F_{\pi}$ is the weak $\pi$-meson decay constant.

Let us rewrite the scalar correlator in the following form
\begin{multline}
\label{ren}
\Pi(Q^2)= 2\sum_{n=0}^{\infty}\frac{G^2(n)m^2(n)}{Q^2+m^2(n)} + D_0 +D_1 Q^2\\
=\sum_{n=0}^{\infty} 2 G^2(n) + D_0 -
Q^2 \left[\sum_{n=0}^{\infty}\frac{2 G^2(n)}{Q^2+m^2(n)} - D_1\right].
\end{multline}
We impose the condition for the constants $D_0$ and $D_1$ that 
all terms at~$Q^2$ and constant contributions (we will not consider low-energy behavior) were cancelled.
It is equivalent to making two subtractions in the point $Q=0$ or just taking two derivatives~\cite{jhep}. 
The same renormalization procedure has to be performed for the pseudoscalar, vector, and axial-vector correlators.

\section{Sum rules}

Let us consider the vector and axial-vector correlators. In the difference
\begin{equation}
\label{vdif}
\Pi^{V}(Q^2)-\Pi^{A}(Q^2)=\mathcal{O}\left(\frac{1}{Q^4}\right),
\end{equation}
both perturbative and gluon condensate contributions are cancelled. It is well known that if the ground states give the main contribution to difference~\eqref{vdif} then expanding in large $Q^2$ one arrives at the following sum rule
\be
\label{2v}
F_{\rho}^2m_{\rho}^2-F_{a_1}^2m_{a_1}^2=0.
\ee
Moreover, if one considers OPE for the divergence of vector and axial-vector currents~\cite{svz,rry} one finds another sum rule
\be
\label{1v}
F_{\rho}^2-F_{a_1}^2-F_{\pi}^2=0,
\ee
The constant $F_{\pi}$ appeared here due to PCAC. S. Weinberg derived these some rules before the invention of large-$N_c$ approach in his paper~\cite{wein} where he then made use of the KSFR relation,
\be
\label{ks}
F_{\rho}^2=2F_{\pi}^2,
\ee
and arrived at his famous formula,
\be 
\label{w} 
m_{a_1}^2=2m_{\rho}^2. 
\ee 
In the framework of large-$N_c$ approach 
the assumption of ground states dominance in~\eqref{vdif} is equivalent to the requirement that the rest of sum is dual to the perturbative QCD continuum. This approximation was then widely used in the so-called Finite Energy Sum Rules (see, e.g., review~\cite{raf}). Within the approximation of infinite number of colours, where we have only resonances, it is more natural to say that this duality means
\be
\label{str}
m_V(n)=m_A(n),\quad n>0.
\ee
Indeed the relation~\eqref{str} is approximately fulfilled in the phenomenology,
the deviations are of order of large-$N_c$ counting (the D-wave vector mesons are not taken into account since they decouple from the sum rules~\cite{jhep}). 

Let us now apply the same logic to the scalar and pseudoscalar
correlators. We will denote the ground state masses and residues as
\be
m_{S,P}(0)\equiv m_{S,P} \qquad G_{S,P}(0)\equiv G_{S,P}.
\ee
The analogue of Eq.~\eqref{2v} is:
\be
\label{2s}
G_{S}^2m_{S}^2-\frac{\langle\bar{q}q\rangle^2}{F_{\pi}^2}=0.
\ee
In this
case we do not have an analogue of the KSFR relation. So we have
to take an additional relation from some other place. Fortunately,
this relation is provided by the sum rules without any appealing
to some "external" formula. Namely, as follows from OPE~\eqref{V} 
and~\eqref{S} for the difference of renormalized correlators
\be
\label{sum}
\Pi^{S,P}(Q^2)-\frac32\Pi^{V,A}(Q^2)=\frac{\alpha_s}{4\pi}
\frac{\langle
G^2\rangle}{Q^2}+\mathcal{O}\left(\frac{1}{Q^4}\right),
\ee
the
contribution of perturbation theory is cancelled. Formula~\eqref{sum} 
contains four relations, three of them being linearly
dependent. Hence we can choose only one of them. We require the
universality for the onset of perturbative continuum in Eq.~\eqref{sum} 
which means in the resonance representation the
following generalization of relation~\eqref{str}
\be
\label{str1}
m_V(n)=m_A(n)=m_S(n)=m_P(n),\quad n>0.
\ee
This relation seems to
be also approximately fulfilled in the phenomenology within the
accuracy of large-$N_c$ approach. In addition, it is supported by
QCD sum rules if we neglect small corrections to the string-like linear
mass spectrum for radial excitations~\cite{qual}. 

Let us consider the $V$-$S$ case. Taking into account only ground states in Eq.~\eqref{sum} and following the standard procedure we obtain the sum rule
\be
\label{2vs}
G_{S}^2m_{S}^2-\frac32F_{\rho}^2m_{\rho}^2=\frac{\alpha_s}{8\pi}\langle G^2\rangle.
\ee

Let us digress for a moment. In the r.h.s. of Eq.~\eqref{2vs} we have the gluon condensate.
It might seem questionable if we may use the physical value for this condensate since it is
impossible to have this value in the "ground state + continuum" approximation or even in the
"several states + continuum" one. We should recall, however, that considering the gluon
condensate contribution has sense only together with taking into account the contribution
due to the perturbation theory (PT); these two contributions are inseparable.
In case of finite number of resonances we can have the physical value for the gluon condensate
only if the contribution of the PT is cancelled like in difference~\eqref{vdif}. It is just what
we have in Eq.~\eqref{sum} in the leading order (LO) of PT. Thus it is quite natural to expect that the
value of gluon condensate in Eq.~\eqref{2vs} is physical.

With this argumentation in mind one can exclude the unknown quantity $G_Sm_S$ from Eqs.~\eqref{2s} and~\eqref{2vs}. Taking into consideration the KSFR relation~\eqref{ks} one obtains the relation
\be
\label{result}
3F_{\pi}^2m_{\rho}^2+\frac18\frac{\alpha_s}{\pi}\langle G^2\rangle=\frac{\langle\bar{q}q\rangle^2}{F_{\pi}^2}.
\ee

Substituting in Eq. \eqref{result} the phenomenological values $F_{\pi}=87$ MeV (the value that it would have in the chiral limit~\cite{gl}), $m_{\rho}=770$ MeV, and $\frac{\alpha_s}{\pi}\langle G^2\rangle=(360\,\text{MeV})^4$ one gets
$\langle\bar{q}q\rangle\approx-(220\,\text{MeV})^3$, i.e. a more or less standard value for the quark condensate. Thus, formula~\eqref{result} gives a right relation between the four important phenomenological quantities.
The result is quite stable against the contribution of PT which is rather large in the scalar sector. For instance, the account of NLO of PT results in a insignificant change of the estimate for the value of quark condensate, $\langle\bar{q}q\rangle\approx-(230\,\text{MeV})^3$.

The quantity in the r.h.s. of Eq.~\eqref{2vs} is numerically by order of magnitude less than the terms in the l.h.s. Consequently within the accuracy of large-$N_c$ approximation the following relation holds
\be
G_{S}^2m_{S}^2\simeq\frac32F_{\rho}^2m_{\rho}^2.
\ee
This means that the amplitude of creation of the ground scalar singlet meson multiplied by the amplitude of annihilation of this meson is approximately equal (up to a factor which supposedly has a kinematic origin) to the same product for the ground vector isovector meson. Since the amplitudes of creation and annihilation are equal this relation can be interpreted as follows: the process of creation and annihilation of the ground scalar meson is, in a sense, approximately dual to the process of creation of two ground vector mesons. The factor 3 in the numerator comes from the summation over the spin degrees of freedom and the factor 2 in the denominator seems to stem from the indistinguishability of two final vector particles.

In the sum rules we deal with the mesons which are genuine quark-antiquark states. According to the modern phenomenology the lightest quark-antiquark scalar isoscalar state is $f_0(1370)$ (see, e.g., note on scalar mesons in~\cite{pdg}). This meson decays mainly into $\rho\rho$-pair while all other mesons with the same quantum numbers dominantly decay into $\pi\pi$-mesons. The duality pointed out above might be a reason for this phenomenon.

\section*{Acknowledgements}

The work was supported by CYT FPA, grant 2004-04582-C02-01, CIRIT GC, grant 2001SGR-00065,
RFBR, grant 05-02-17477, and by Ministry of Education and Science of Spain.

\end{document}